\documentclass[a4paper,11pt]{article}
\usepackage{pos}
\usepackage[english]{babel}
\usepackage{amsmath,graphicx,bbm,xcolor}

\newcommand{\mathd}{\mathrm{d}}
\newcommand{\nosymbol}{}
\newcommand{\tmcolor}[2]{{\color{#1}{#2}}}
\newcommand{\tmop}[1]{\ensuremath{\operatorname{#1}}}
\newcommand{\tmscript}[1]{\text{\scriptsize{$#1$}}}

\newcommand{\tmtextit}[1]{\text{{\itshape{#1}}}}

\title{Real time dynamics of a semiclassical gravitational collapse of a scalar quantum field}

\author[a,b]{Jana N. Guenther}
\author*[a]{Christian Hoelbling}
\author[a,b]{Lukas Varnhorst}

\affiliation[a]{Department of Physics, University of Wuppertal,
  Gau{\ss}strasse 20, D-42119 Wuppertal, Germany}

\affiliation[b]{Aix Marseille Univ., Universit\'e de Toulon, CNRS, CPT,
  Marseille, France}

\emailAdd{jguenther@uni-wuppertal.de}
\emailAdd{hch@uni-wuppertal.de}
\emailAdd{varnhorst@uni-wuppertal.de}

\abstract{We present a new formalism for numerically treating the
  semiclassical gravitational collapse of a scalar quantum field in
  the radially symmetric case. Our formalism is time reversal
  invariant and the evolution of the scalar fields is unitary. We
  present some first results in the angular momentum $l=0$
  approximation for an initially coherent state of a massless field
  and briefly discuss future prospects.}

\FullConference{%
 The 38th International Symposium on Lattice Field Theory, LATTICE2021
  26th-30th July, 2021
  Zoom/Gather@Massachusetts Institute of Technology
}

\begin{document}

\maketitle

\section{Introduction}

The formation of black holes via gravitational collapse and their subsequent
evaporation via Hawking radiation {\cite{Hawking:1974sw}} is a process whose
dynamics is not yet satisfactorily understood. Specifically, it is not clear
whether it can be described by a unitary time evolution or there is an
inherent loss of information (see e.g. {\cite{Mathur:2009hf,Harlow:2014yka}}
for recent reviews on this topic). In this note, we describe a formalism whose
ultimate aim it is to provide some insight into this question numerically, in
a simple, semiclassical setup. We investigate a massless, scalar quantum
field, coupled semiclassically to classical gravity via a dynamical metric, but otherwise
free. For simplicity, our setup is spherically symmetric and the background
metric is assumed to be flat and without singularities in the absence of the
field. Our aim is to compute the real time evolution of the scalar field and
the associated metric, starting from an inmoving, coherent state. We show
that, in principle, the time evolution of the system can be carried out
numerically and present some results in the $l = 0$ approximation. Finally, we
briefly touch upon issues of vacuum subtraction of the higher angular momentum
modes. More detailed results for the $l = 0$ approximation can be found in
{\cite{Guenther:2020kro}} and a different formalism for treating this problem
is presented in {\cite{Berczi:2020nqy,Berczi:2021hdh}}.

\section{Derivation of the formalism}

We work with the spherically symmetric metric
{\cite{Christodoulou:1986zr,Christodoulou:1987vv,Goldwirth:1987nu,Choptuik:1992jv}}
\[ \mathd \tau^2 = \alpha^2 (t, r) \mathd t^2 - a^2 (t, r) \mathd r^2 - r^2
   \mathd \Omega^2 \]
and the action
\[ S_{\pm} = \int \mathd^4 x \left( \sum_{i = 1}^{N_f} \frac{(-
   1)^{P_i}}{2}  \left( g^{\alpha \beta} (x) 
   {\overline{\phi}^{(i)}}^{\dag}_{, \alpha} (x) \overline{\phi}^{(i)} (x)_{,
   \beta} - M_i^2 {\overline{\phi}^{(i)}}^{\dag} (x) \overline{\phi}^{(i)} (x)
   \right) - \frac{1}{16 \pi} R (t, r) \right) \]
which allows us to in principle include fields of different masses $M_i$. Any
number of these fields can be Pauli-Villars regulators (as suggested in
{\cite{Berczi:2020nqy,Berczi:2021hdh}}) if we chose $P_i = 1$
(otherwise $P_i = 0$). We decompose
our scalar fields into spherical harmonics
\[ \overline{\phi} (t, r, \theta, \varphi) = \sum_{l = 0}^{\infty} \sum_{m = -
   l}^l \overline{\phi}_{l m} (t, r) Y_{l m} (\theta, \varphi) \]
and rescale them as
\[ \phi_{l m} = \overline{\phi}_{l m} r \sqrt{\frac{a_0}{2 \alpha_0}} \]
where $a_0$ and $\alpha_0$ are the metric parameters at the initial time
$t_0$. This results in a Hamiltonian density that is diagonal in the scalar
field components
\[ \mathcal{H}_{l m}^{(i)} = \Pi^{(i)}_{l m} {A \Pi^{(i)}_{l m}}^{\dag} {+
   \phi^{(i)}}^{\dag}_{l m} A^{- \frac{1}{2}} K^{(i)}_l \phi^{(i)}_{l m} \]
where the real, symmetric operators $K^{(i)}_l$ are given by
\begin{equation}
  K_l^{(i)} = q^T q + \alpha^2 \left( \frac{l (l + 1)}{r^2} + M^2 \right)
  \qquad q = r \sqrt{\frac{\alpha}{a}} \partial_r \frac{1}{r} 
  \sqrt{\frac{\alpha}{a}} \label{Kqdef}
\end{equation}
and
\[ A = \frac{a_0 \alpha}{\alpha_0^{\nosymbol} a} \]
Note that at the initial time $t_0$ the factor $A = 1$. We proceed by
canonical quantization of the fields $\phi^{(i)}_{l m}$. Parameterising our
fields in the initial time Fock basis with annihilation operators $b^{(i)}_{l m \pm}$, we can define time evolution
coefficient matrices $u_l^{(i)} (t)$ and $v_l^{(i)} (t)$ via
\[ \phi^{(i)}_{l m} = \frac{{u^{(i)}_l}^{\dag} (t) {b^{(i)}_{l m +}}^{\dag} +
   u_l^{(i) T} (t) b^{(i)}_{l m -}}{\sqrt{2}} \qquad \Pi^{(i)
   }_{l m} = \frac{b^{(i)}_{l m +} v^{(i)}_l (t) {- b^{(i)}_{l m
   -}}^{\dag} v_l^{(i) \ast} (t)}{\sqrt{2}} \]
Diagonalizing the Hamiltonian at the initial time $t_0$ via
\begin{equation}
  K^{(i)}_l = V_l^{(i)} \omega^{(i) 2}_l V_l^{(i) T} \label{evdec}
\end{equation}
we find the initial condition for the evolution coefficients to be
\begin{equation}
  u_l^{(i)} (t_0) = \frac{1}{\sqrt{\omega_l^{(i)}}} V_l^{(i) T} \qquad
  v_l^{(i)} (t_0) = \sqrt{\omega_l^{(i)}} V_l^{(i) T} \label{initc}
\end{equation}
and their Heisenberg picture time evolution given by
\begin{equation}
  \dot{u}^{(i)}_l = - i v_l^{(i)} A \qquad \dot{v}_l^{(i)} = - i u_l^{(i)}
  A^{- \frac{1}{2}} K^{(i)}_l A^{- \frac{1}{2}} \label{tev}
\end{equation}
Note that for a fixed metric, this time evolution is a Bogolyubov
transformation and thus fulfills the identities
\[ \tmop{Re} \left( {u_l^{(i)}}^{\dag} v_l^{(i)} \right) = \tmop{Re} \left(
   {v_l^{(i)}}^{\dag} u_l^{(i)} \right) =\mathbbm{1} \qquad \tmop{Im} \left(
   {u_l^{(i)}}^{\dag} u_l^{(i)} \right) = \tmop{Im} \left( {v_l^{(i)}}^{\dag}
   v_l^{(i)} \right) = 0 \]
With the Heisenberg time evolution settled, we next need to define an initial
state. For this purpose, we first define an annihilation operator
\[ b = b^{(0)}_{00 + k} f_{+ k} + b^{(0)}_{00 - k} f^{\ast}_{- k} \]
where $k$ is an explicit momentum index which is summed over. Note that we
only use $l = 0$ operators to maintain radial symmetry. Also note,
that the operator $b$ only excites a single field $i = 0$. We use this
annihilation operator to produce an initially coherent state of unit norm
\[ | \lambda \rangle = e^{- \frac{| \lambda |^2}{2}} \sum_{n = 0}^{\infty}
   \frac{\lambda^n}{n!} {b^{\dag}}^n | 0 \rangle \]
The coefficients $f_{\pm k}$ encode the radial shape of our state. Using the
shorthand notation
\[ l_+ = \lambda^{\ast} f_+ \qquad l_- = \lambda f_- \]
and
\[ \begin{array}{rll}
     l_u & = & {u_0^{(0)}}^{\dag} l_+ - {u_0^{(0)}}^T l_-\\
     l_v & = & -i({v_0^{(0)}}^{\dag} l_+ + {v_0^{(0)}}^T l_-)
   \end{array} \]
we can take the expectation values of the elements of the energy momentum
tensor in these states and write down the field equations. After some algebra
this results in
\begin{equation}
  \begin{array}{rll}
    \ln' (a \alpha) & = & h_r\\
    &  & \\
    \left( \ln' \left( \frac{a}{\alpha} \right) - \frac{1 - a^2}{r} \right)
    \frac{1}{a^2} & = & m_r\\
    &  & \\
    \frac{\dot{a}}{\alpha} & = & p_r
  \end{array} \label{feq}
\end{equation}
where the densities \ and $p_r$ at the radial coordinate $r$ are given as
\begin{equation}
  \begin{array}{rll}
    h_r & = & \frac{a^0}{r \alpha^0}  (| (q^0 l_u)_r |^2 + | l_{v r} |^2) +
    h^0_r\\
    &  & \\
    h^0_r & = & \frac{a^0}{r \alpha^0}  \left( \sum_i (- 1)^{P_i} \sum_{l =
    0}^{\infty} (2 l + 1) \left( \left( {v_l^{(i)}}^{\dag} v^{(i)}_l
    \right)_{r r} + \left( q^0 {u^{(i)}_l}^{\dag} u^{(i)}_l q^{0 T} \right)_{r
    r} \right) \right)\\
    &  & \\
    m_r & = & \frac{\alpha_0}{r a_0} \sum_i (- 1)^{P_i} \sum_{l = 0}^{\infty}
    (2 l + 1)  \left( \frac{l (l + 1)}{r^2} + M_i^2 \right) \left(
    {u^{(i)}_l}^{\dag} u^{(i)}_l \right)_{r r}\\
    &  & \\
    p_r & = & \frac{a_0}{r \alpha_0^{\nosymbol}} \left( \tmop{Im} (l_{v r} 
    (q^0 l_u)_r) + \sum_i (- 1)^{P_i}  \sum_{l = 0}^{\infty} (2 l + 1)
    \tmop{Im} \left( q^0 {u^{(i)}_l}^{\dag} v^{(i)}_l \right)_{r r} \right)
  \end{array} \label{densities}
\end{equation}
with the operator
\[ q^0 = q (t_0) = r \sqrt{\frac{\alpha_0}{a_0}} \partial_r \frac{1}{r} 
   \sqrt{\frac{\alpha_0}{a_0}} \]
Note that none of the right hand sides in (\ref{densities}) contain a
reference to the current metric, which allows for a convenient radial
integration of the first two equations in (\ref{feq}). In fact, defining
\[ \hat{\alpha} = \alpha a \qquad d = \frac{r}{a^2} \]
they decouple and can be written as
\[ \begin{array}{rll}
     \ln' (\hat{\alpha}) & = & h_r\\
     &  & \\
     d' + d h_r & = & 1 - r m_r
   \end{array} \]

\section{Discretization and regularization}

We discretize our system with $N_r$ radial shells of equal coordinate distance
$\Delta = r_i - r_{i - 1}$ (more general discretizations might be considered).
One step beyond the innermost shell, we have the boundary condition $d_0 = 0$,
which implies that for $r_0 = 0$ there is no central singularity. The case
$r_0 > 0$, corresponding to a horizon at $r_0$, is not considered here.
Outside the outermost shell, we assume a Schwarzschild metric, so our second
radial boundary condition is $\hat{\alpha}_{N_r} = 1$. Assuming that the
densities $h_r$ and $m_r$ are concentrated in $\delta$-shells at the radial
coordinates, we find that radial integration can be accomplished by
\begin{equation}
  \begin{array}{lll}
    \hat{\alpha}_i & = & e^{- \sum_{j = i + 1}^{N_r} h_j}\\
    d_i & = & e^{- h_i^0} (d_{i - 1} + \Delta) + \frac{e^{- h_i^0} - 1}{h_i^0}
    m_i r_i
  \end{array} \label{radintd}
\end{equation}
It is clear however, that we can not use the densities $h_r$ and $m_r$ as
given in (\ref{densities}), since they contain divergent vacuum
contributions.\footnote{Note however, that the vacuum contributions of $p_r$
vanish at the initial time $t_0$. Although this is not pursued here, it might
help in finding an alternative integration scheme that requires less
regularization.} A first step towards regularization, which turns out to be
sufficient in the $l = 0$ approximation, is the normal ordering of the
operators. In the simplest case, to which we will restrict our attention here,
the normal ordering is performed once at the initial time $t_0$ with no
further corrections, resulting in
\[ : h^0_r : (t) = h^0_r (t) - h^0_r (t_0) \hspace{3em} : m_r : (t) = m_r (t)
   - m_r (t_0) \]
We also need to discretize the derivative operator $q$ in (\ref{Kqdef}),
which we will do here with the simple forward difference operator and an inner
and outer shell truncation, corresponding to reflecting boundary conditions.
For more generic choices on all of the above, we refer to
{\cite{Guenther:2020kro}}, where different variations are worked out in detail
for the $l = 0$ approximation. For the temporal evolution of the field
component matrix, we adopt a time reversible integration scheme that respects
the Bogolyubov identities. This is achieved by alternating between an implicit
and explicit time step
\[ \left( \begin{array}{lll}
     u_{l, t + \Delta t}^{(i)} &  & v_{l, t + \Delta t}^{(i)}
   \end{array} \right) = \left( \begin{array}{lll}
     u_{l, t}^{(i)} &  & v_{l, t}^{(i)}
   \end{array} \right) e^{\tmscript{- i \left(\begin{array}{cc}
     0 & \frac{1}{\sqrt{A}} \tmcolor{brown}{} K_l^{(i)} \frac{1}{\sqrt{A}}\\
     A & 0
   \end{array}\right) \Delta t}} \]
which results from a straightforward integration of (\ref{tev}). For the
numerical implementation, we perform a diagonalization (\ref{evdec}) of the
operator $K_l^{(i)}$ from (\ref{Kqdef}), which (leaving out the obvious
indices $l$ and $(i)$ for compactness) results in
\begin{equation}
  \begin{array}{rll}
    u_{t + \Delta t} & = & \left( u_t \frac{1}{\sqrt{A}} V \cos (\omega \Delta
    t) - i  v_t \sqrt{A} V \frac{1}{\omega} \sin (\omega \Delta t) \right) V^T
    \sqrt{A}\\
    v_{t + \Delta t} & = & \left( v_t \sqrt{A} V \cos (\omega \Delta t) - i 
    u_t \frac{1}{\sqrt{A}} V \omega \sin (\omega \Delta t) \right) V^T
    \frac{1}{\sqrt{A}}
  \end{array} \label{uvexp}
\end{equation}
In the implicit step, we iterate between the update (\ref{uvexp}) and a radial
integration of the metric (\ref{radintd}). The necessary diagonalizations in
this step make up the bulk of the computational expense.

Finally, we want to specify our choice for the initial wave packet. It turns
out to be advantageous to take a window function, which vanishes identically
outside a specified range. In the results presented, we opt for a Nuttall
window {\cite{1163506}}
\[ f_{R, \sigma, \lambda} (r) = \lambda \left( \sum_{k = 0}^3 a_k \cos \left(
   k \frac{\pi (r - R + \sigma)}{\sigma} \right) \right)^2 \]
with the coefficients
\[ a_1 = - 0.487396 \qquad a_2 = 0.144232 \qquad a_3 = - 0.012604 \qquad a_0 =
   - a_1 - a_2 - a_3 \]
and an amplitude $\lambda$, which is peaked at $R$ and vanishes for $| R - r |
\geqslant \sigma$. The metric is then initialized by setting
\[ h_i = f_{R, \sigma, \lambda} (r_i) \qquad m_i = 0 \]
and performing a radial integration (\ref{radintd}). To initialize the state,
we then chose an inmoving wave packet $\kappa_i = 1$ and set
\[ l_{u i} = \tmop{sign} (\kappa_i) \sqrt{h_i \left( 1 \mp \sqrt{1 -
   \kappa_i^2} \right)} \hspace{4em} l_{v i} = \sqrt{h_i \left( 1 \pm \sqrt{1
   - \kappa_i^2} \right)} \]
We then compute the initial $K^{(i)}_l$ according to (\ref{Kqdef}) and perform
their eigenmode decomposition (\ref{evdec}), which allows us to construct the
initial evolution coefficient matrices (\ref{initc}), so we can finally
initialize the state vector components as
\[ l_{\pm} = \frac{1}{2 \omega^{(0)}_0} (\pm u^{(0)}_0 q^{0 T} l_u + i
  v^{(0)}_0 l_v) \]

\section{Some results in the $l = 0$ approximation}

\begin{figure}[h]
  \resizebox{13.99cm}{!}{\includegraphics{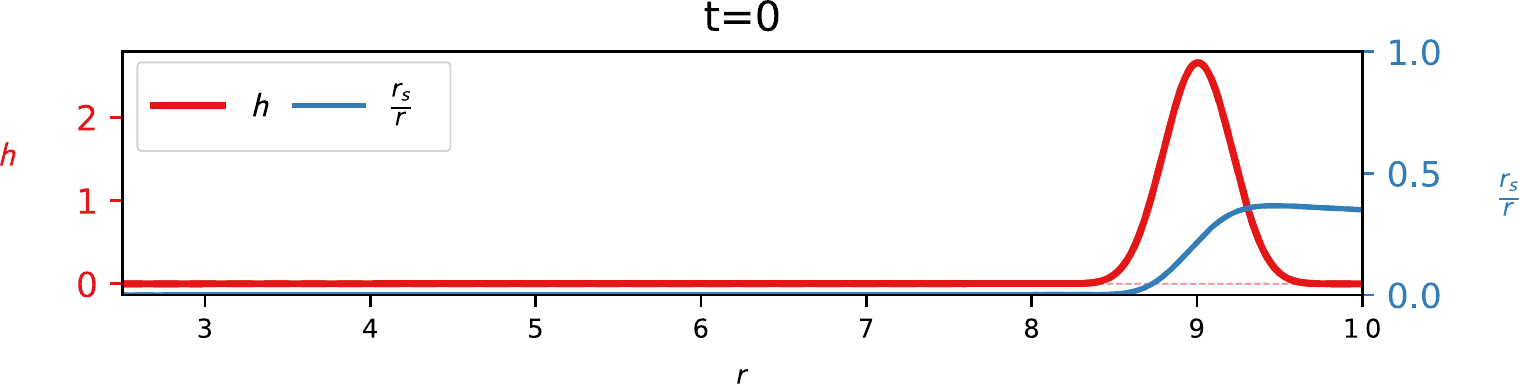}}
  \caption{\label{initb}The initial normal ordered Hamiltonian density $h =
  \frac{\alpha}{2 a} h_r$ and the ratio of the local Schwarzschild radius $r_s
  = r - d$ to the local radius.}
\end{figure}

For a first test of our formalism, we restrict ourselves to the $l = 0$
approximation. In that case, we do not require any further regularization
beyond the normal ordering. In order to make the vacuum effects more
prominent, we choose to study two massless fields $N_f = 2$ with only the
first one carrying a classical component. The results presented here
correspond to our standard discretization with $N_r = 800$ radial
discretization points and a radial extent from $r_0 = 0$ to $r_{\max} = 10$, a
time step $\Delta t = 0.004$, an exterior Schwarzschild metric with a total
$r_s = 3.5$ and an inital bump with a width $\sigma = 0.5$ around $R = 9$. The
initial state at $t = 0$ is depicted in fig.~\ref{initb}.

\begin{figure}[h]
  \resizebox{13.99cm}{!}{\includegraphics{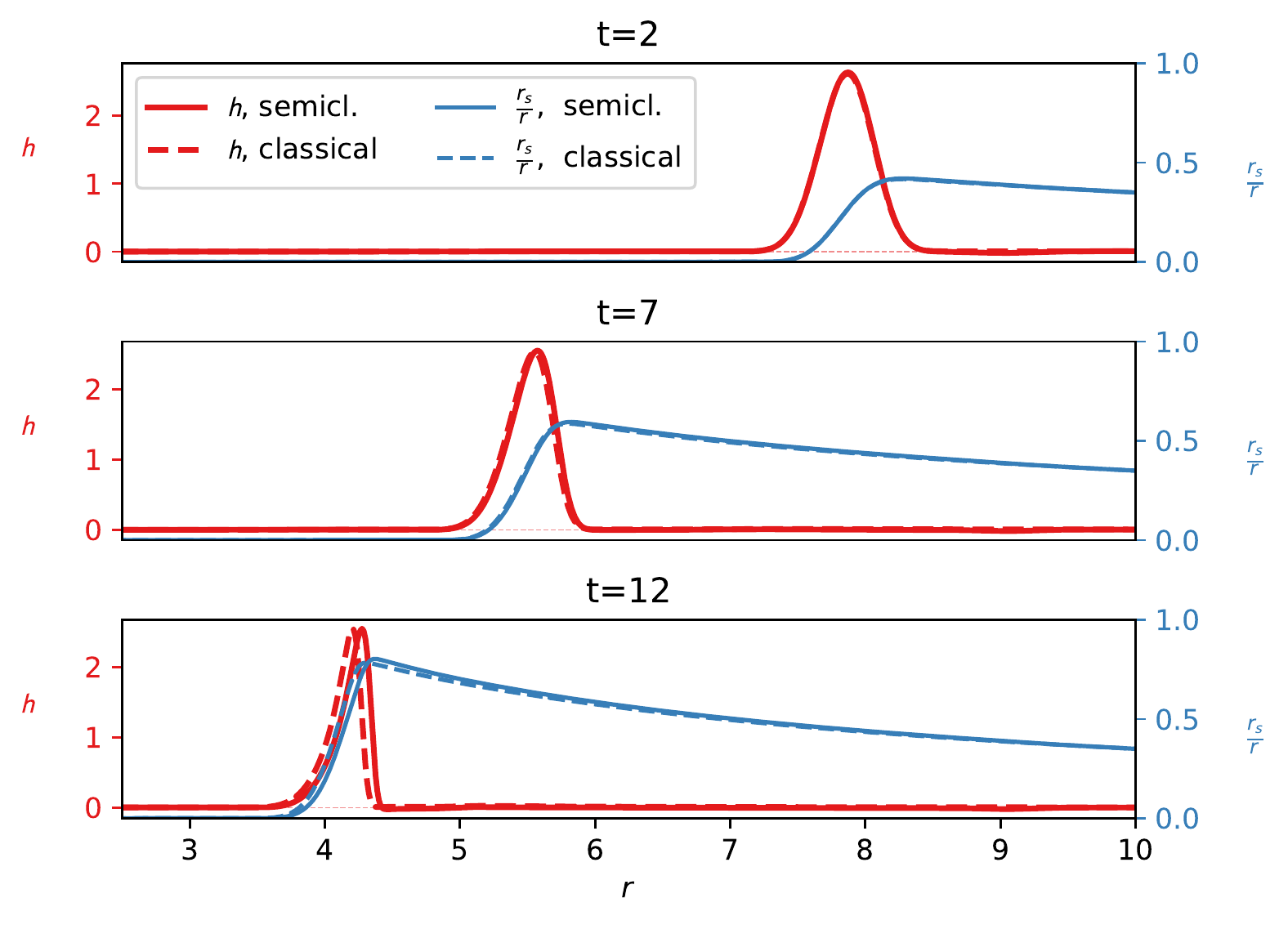}}
  \caption{\label{ev1}Time evolution of the state from fig.~\ref{initb}.}
\end{figure}

Performing the time evolution of our system in both the classical and the
semiclassical cases, we observe the onset of horizon formation (see fig.
\ref{ev1}), which seems to happen at larger radial coordinates in the
semiclassical case.

\begin{figure}[h]
  \resizebox{13.99cm}{!}{\includegraphics{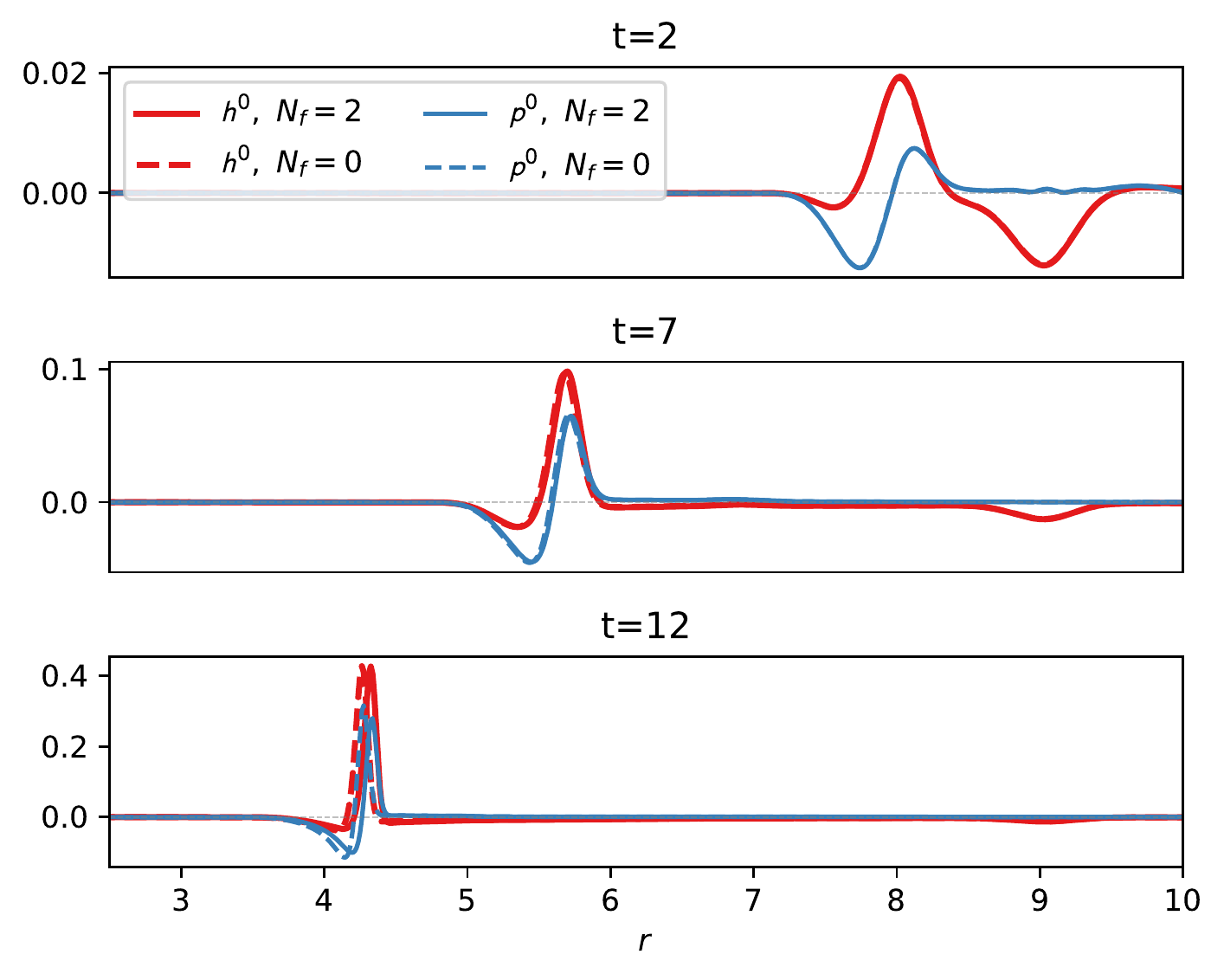}}
  \caption{\label{vacev}The vacuum part of the Hamiltonian density for the
  Hamiltonian density $h^0$ and the radial momentum density $p^0 =
  \frac{\alpha}{2 a} p^0_r$ for three different times in the evolution. Note
  the different scales of the $y$-axis in the three plots. The static dip
  around $r = 9$ is a remnant of our initial state vacuum definition.}
\end{figure}

In order to highlight the difference between the classical and semiclassical
cases in the $l = 0$ approximation, we can plot the vacuum terms alone. We can
in fact do this for the classical case, where we can measure the vacuum term
even though it does not contribute to the time evolution. In fig.~\ref{vacev}
we have plotted the corresponding terms, providing evidence that within our
approximation the vacuum in both the semiclassical and the classical time
evolution tend to enhance the peak energy density.

\begin{figure}[h]
  \resizebox{13.99cm}{!}{\includegraphics{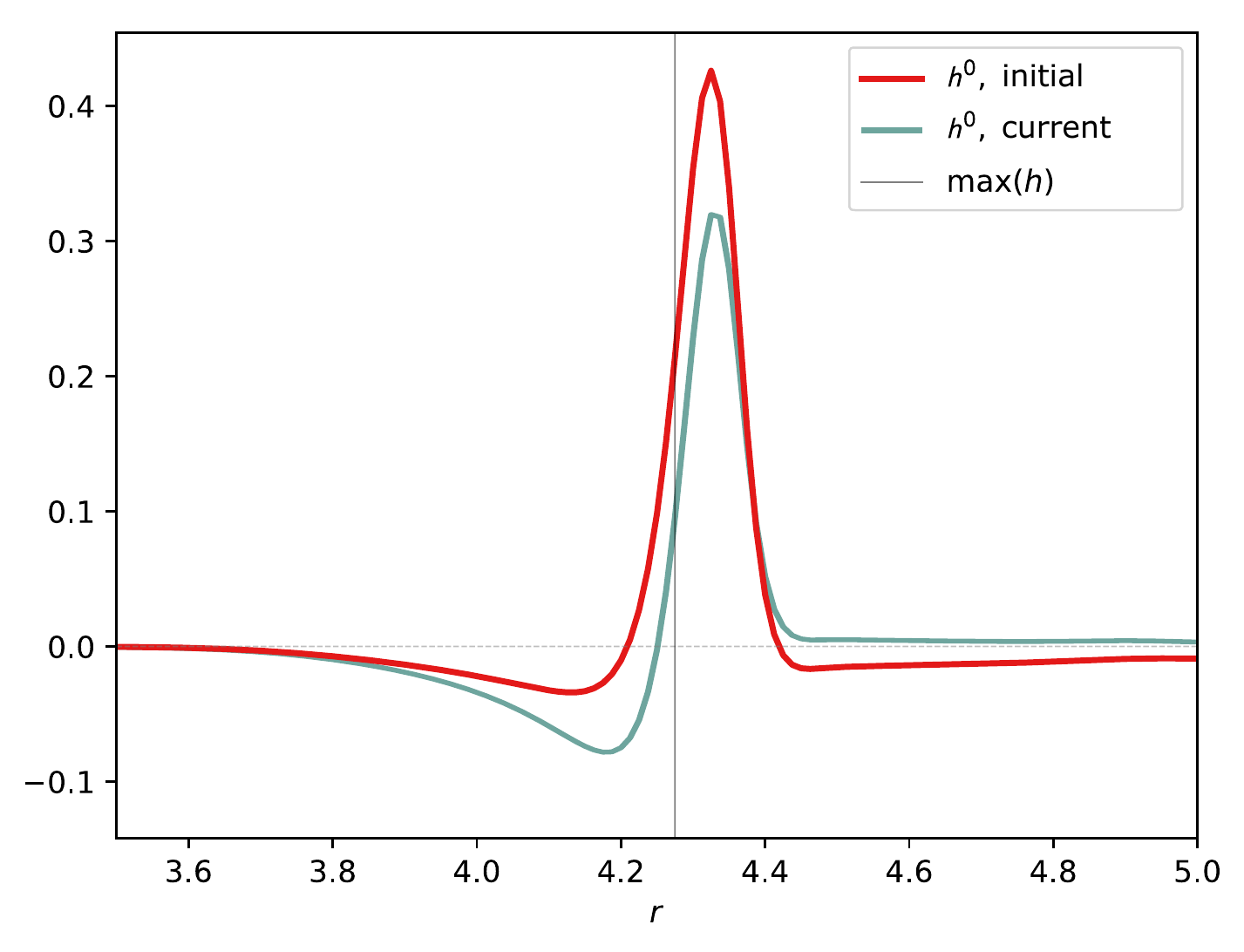}}
  \caption{\label{vacomp}Comparison of the vacuum energy contribution at $t =
  12$ as seen with respect to the either the initial or current metric vacuum
  energy subtraction. The vertical line corresponds to the maximum of the
  total Hamiltonian density at this time.}
\end{figure}

In fig.~\ref{vacomp}, we zoom in on the vacuum contribution of the
semiclassical time evolution at $t = 12$. In addition to the vacuum term with
respect to the initial metric, we have also plotted the vacuum term with
respect to the current metric in the time evolution. The latter is obtained by
not subtracting $h_0 (t)$, but instead by computing a new $h_0 (t_0)$
according to (\ref{densities}) with the evolution coefficients reinitialised
at $t_0 = t$ as in (\ref{initc}). One can see, that with respect to the
initial state metric, the vacuum term enhances the peak region while
depressing both outer and inner flanks, whereas with respect to the current
state metric, the vacuum contribution is negative inside the peak region and
positive just outside.

\section{Conclusions and outlook}

We are investigating the semiclassical gravitational collapse of a massless
scalar field. While the formalism seems to work in principle and yields
interesting first results in the $l = 0$ approximation, there still is a lot
of conceptual work to be done. Most importantly, we want to go beyond the $l =
0$ approximation, which requires additional regularization. We are currently
in the process of investigating different approaches, including static mode
subtraction, point splitting and the introduction of Pauli-Villars fields as
suggested in {\cite{Berczi:2020nqy,Berczi:2021hdh}}. There are also a number
of technical challenges, such as instabilities in the eigenmode decomposition
near the origin or the proper treatment of high frequency modes close to a
forming horizon, which have to be overcome before more physical conclusions
about the behaviour of the system can be drawn.


\begin{thebibliography}{10}
  \bibitem[1]{Hawking:1974sw}S.W.~Hawking. {\newblock}Particle Creation by
  Black Holes. {\newblock}\tmtextit{Commun. Math. Phys.}, 43:199--220, 1975.
  {\newblock}[Erratum: Commun.Math.Phys. 46, 206 (1976)].{\newblock}
  
  \bibitem[2]{Harlow:2014yka}Daniel Harlow. {\newblock}Jerusalem Lectures on
  Black Holes and Quantum Information. {\newblock}\tmtextit{Rev. Mod. Phys.},
  88:15002, 2016.{\newblock}
  
  \bibitem[3]{Mathur:2009hf}Samir~D.~Mathur. {\newblock}The Information
  paradox: A Pedagogical introduction. {\newblock}\tmtextit{Class. Quant.
  Grav.}, 26:224001, 2009.{\newblock}
  
\bibitem[4]{Guenther:2020kro}Jana~N.~Guenther, Christian Hoelbling, and 
  Lukas Varnhorst. {\newblock}Semiclassical gravitational collapse of a
  radially symmetric massless scalar quantum field. {\newblock}10
  2020.{\newblock}
  
  \bibitem[5]{Berczi:2021hdh}Benjamin Berczi, Paul~M.~Saffin, and  Shuang-Yong
  Zhou. {\newblock}Gravitational collapse of quantum fields and Choptuik
  scaling. {\newblock}11 2021.{\newblock}
  
  \bibitem[6]{Berczi:2020nqy}Benjamin Berczi, Paul~M.~Saffin, and  Shuang-Yong
  Zhou. {\newblock}Gravitational collapse with quantum fields.
  {\newblock}\tmtextit{Phys. Rev. D}, 104(4):0, 2021.{\newblock}
  
  \bibitem[7]{Choptuik:1992jv}Matthew~W.~Choptuik. {\newblock}Universality and
  scaling in gravitational collapse of a massless scalar field.
  {\newblock}\tmtextit{Phys. Rev. Lett.}, 70:9--12, 1993.{\newblock}
  
  \bibitem[8]{Christodoulou:1986zr}D.~Christodoulou. {\newblock}The Problem of
  a Selfgravitating Scalar Field. {\newblock}\tmtextit{Commun. Math. Phys.},
  105:337--361, 1986.{\newblock}
  
  \bibitem[9]{Christodoulou:1987vv}D.~Christodoulou. {\newblock}A Mathematical
  Theory of Gravitational Collapse. {\newblock}\tmtextit{Commun. Math. Phys.},
  109:613--647, 1987.{\newblock}
  
  \bibitem[10]{Goldwirth:1987nu}Dalia~S.~Goldwirth  and  Tsvi Piran.
  {\newblock}Gravitational Collapse of Massless Scalar Field and Cosmic
  Censorship. {\newblock}\tmtextit{Phys. Rev. D}, 36:3575, 1987.{\newblock}

\bibitem[11]{1163506}A.~Nuttall. {\newblock}Some windows with very good
  sidelobe behavior. {\newblock}\tmtextit{IEEE Transactions on Acoustics,
  Speech, and Signal Processing}, 29(1):84--91, 1981.{\newblock}
\end{thebibliography}
\end{document}